\documentclass[preprint,11pt,authoryear,nonatbib]{elsarticle}

\usepackage[a4paper,margin=2.5cm]{geometry}
\usepackage{setspace}
\usepackage[utf8]{inputenc}
\usepackage[T1]{fontenc}

\usepackage{algorithm}
\usepackage{algpseudocode}
\usepackage{amsfonts}
\usepackage{amssymb}
\usepackage{amsmath}
\usepackage{float}
\usepackage{graphicx}
\usepackage{subcaption}
\usepackage{booktabs}
\usepackage{mathtools}
\usepackage{enumerate}
\usepackage{xcolor}
\usepackage[american]{babel}
\usepackage{csquotes}
\usepackage[style=apa,backend=biber,natbib=true]{biblatex}
\usepackage{xurl}

\DeclareMathOperator*{\argmin}{arg\,min}

\journal{European Journal of Operational Research}

\begin{document}

\begin{frontmatter}

\title{A Functional Representation of Credit Behavior for Probability of Default Modeling}

\author[nyk,aaumath]{Jonas Brunholm\corref{cor1}}
\ead{jbjens@math.aau.dk}
\author[adv]{Bjarne H\o jgaard}
\author[aaucs]{Thomas Dyhre Nielsen}
\author[aaumath]{Orimar Sauri}

\cortext[cor1]{Corresponding author.}

\affiliation[nyk]{organization={Nykredit},
                  addressline={Skelagervej 15},
                  postcode={9000},
                  city={Aalborg},
                  country={Denmark}}
\affiliation[aaumath]{organization={Department of Mathematical Sciences, Aalborg University},
                  addressline={Thomas Manns Vej 23},
                  postcode={9220},
                  city={Aalborg {\O}},
                  country={Denmark}}
\affiliation[adv]{organization={Advisense},
                  addressline={Dronningens Tv{\ae}rgade 7},
                  postcode={1302},
                  city={Copenhagen},
                  country={Denmark}}
\affiliation[aaucs]{organization={Department of Computer Science, Aalborg University},
                  addressline={Selma Lagerl{\o}fs Vej 300},
                  postcode={9220},
                  city={Aalborg {\O}},
                  country={Denmark}}

\begin{abstract}
This paper proposes a framework for modeling probability of default via functional data analysis. By representing a series of credit variables as functions, we investigate whether intra-monthly information improves default predictability in linear models.
We further show how a range of widely used variables, among them available funds, utilization rate, and overdraft, can all be derived from three quantities observed over time. Namely, 1) the type of each account, 2) the balance of the account, and 3) the size of the credit limit on the account. Retaining these quantities as continuous-time processes, rather than reducing them to monthly aggregated values, yields a continuous faithful representation of the borrower. When analyzing these three processes, we discovered that recurring events associated with the ordinal position among banking days created strong cyclical patterns. We therefore develop a relative time framework that aligns the recurring events across borrowers, ensuring that borrowers possess the same cyclical pattern, regardless of real time.
We assess the framework using functional logistic regression. This approach accommodates the continuous representation while remaining closely related to a logistic regression model commonly used in credit risk practice, due to strict regulatory constraints. We show that, when equipped with an effective functional representation of transactional trajectories, the proposed model attains predictive performance on par with XGBoost while consistently outperforming logistic regression. Importantly, the model balances predictive performance of a machine learning model with the interpretability of linear default models, potentially enabling financial institutions to use the model, even under strict regulation.
\end{abstract}


\end{frontmatter}

\section{Introduction}\label{sec:introduction}
The Probability of Default (PD) is a central quantity in credit risk management. It informs lenders of the probability that a borrower fails to meet their financial obligations within a given horizon. It is used when lenders price loans, provision for losses, and allocate regulatory capital under the Internal Ratings Based (IRB) framework. Because of this, improvements in ranking capabilities of PD estimates have been shown to improve profitability for financial institutions \citep{GIUDICI2026100122}. PD modeling has therefore received considerable attention from both practitioners and researchers.

The industry standard for PD modeling remains logistic regression, which is simple, transparent, easy to justify to regulators, and continues to serve as the baseline in empirical comparisons \citep{LESSMANN2015124}. Around the logistic regression benchmark, a body of work shows that machine learning (ML) models such as neural networks and XGBoost models can improve predictive accuracy for borrower-level default risk by capturing nonlinearities in borrower characteristics \citep{KHANDANI20102767,XIA2017225,Addo_risks6020038}. Furthermore, the field has increasingly moved toward alternative data sources and more flexible estimators \citep{Longitudinal_Systematic_Review_data,ZHANG2024121484}. This shift has, in turn, raised concerns about interpretability, fairness, and governance \citep{KOZODOI20221083}, prompting a parallel literature on explainable AI in credit risk that proposes tools such as SHAP values and other local explanation methods \citep{CHACON2026101424}. Alongside this effort are the legal and ethical analyses of discrimination and opacity when big-data and ML techniques are used for scoring \citep{hurley2016credit}.

A common characteristic across this literature is the treatment of the data itself. Whether the downstream model is a logistic regression or a neural network, the data is generally taken as a fixed input: transactional histories are first reduced to a fixed set of summary variables, \textit{typically averages over months or quarters}, or the value at the time of observation, and then passed to the model \citep{OgundimuEmmanuelO.2019Podp,WANG2024114084}. This aggregation step is usually regarded as a necessary preprocessing choice, given the irregular and noisy nature of raw transactional data, and it is rarely examined in its own right. It is standard practice in the Danish financial industry, and it is also reflected in the publicly available benchmark datasets that much of the research relies on, such as the Taiwan dataset described in \citet{OgundimuEmmanuelO.2019Podp} and the German and GMSC datasets described in \citet{KOZODOI20221083}. Much of the effort thus goes into extracting more information from the data through more flexible models, while the aggregated representation of the data is held fixed. Aggregation is, however, not without cost: by summarizing a month into a single number, it discards the within-month dynamics of how a borrower behaves.

In this work we exploit that the data underlying most credit risk models can be created using three central processes, namely, 1) the type of each account, 2) the balance of the account, and 3) the size of the credit limit on the account. Thus, the well known credit variables are not separate pieces of information, but different reductions of the same three underlying processes. Rather than commit to a fixed set of end-of-month figures in advance, we keep each quantity as a process over time. Using daily observations of the three processes, we show that retail borrower variables are cyclical, tied to ordinal positions during the month, namely, salary payments on the last banking day and increased spending during the first banking days.

With this information, we construct a relative time grid over banking days, ensuring that the ordinal positions occur simultaneously on a common time grid.
The aligned processes contain both different observational length and different within-month observational frequencies, due to using a banking day grid when aligning. These challenges motivate the use of functional data analysis (FDA), which treats each observed history as a single realization of an underlying random function rather than as a fixed-length vector of features \citep{FDA_Book_Ramsay_2005}. Using the functional representations of credit variables, we estimate defaults via functional logistic regression \citep{goldsmith2011penalized}. We chose this approach because the processes have an unequal amount of observations, both across borrowers and within months. Furthermore, the functional logistic regression belongs to the class of linear models, allowing for potential use in a regulatory setting.

To the best of our knowledge, this is the first paper to combine functional data analysis with intra-monthly observations to model borrower defaults. Related studies have exploited temporal information using alternative approaches. In particular, \citet{kvamme2018predicting} used daily account balances as raw inputs in a convolutional neural network, while \citet{Alaraj2021} applied an     LSTM neural network to monthly credit card information to predict missed payments. In contrast, our approach retains daily information and its temporal structure while preserving interpretability.

To assess the performance of our proposed methodology, we compare our functional logistic regression model with the logistic regression and an XGBoost model, which serves as a top machine learning benchmark \citep{GUNNARSSON2021292}. We find that the resulting model coefficient functions are consistent with credit intuition and with the coefficients of the standard logistic model. Moreover, the functional model outperforms the standard logistic regression while attaining performance comparable to an XGBoost benchmark on a range of loss metrics, including the AUC.

We note that the representation developed here is not specific to PD modeling. Since the three quantities underlie credit behavior more generally, the same continuous-process view may also be relevant for behavioral monitoring, and other credit risk quantities. We restrict attention to PD modeling in this paper and leave these wider applications for future work.

We can summarize our main contributions in three parts: First, we introduce a framework that represents a borrower through three continuous-time processes, account type, balance, and outstanding credit, from which the established credit behavior variables are constructed. Second, we show how this description can be used directly for credit risk modeling, developing a functional logistic regression for the probability of default as a concrete demonstration. Third, we find empirically, both in-sample and on an out-of-time sample, that this approach significantly outperforms the standard logistic regression and matches an XGBoost benchmark, while remaining a transparent and auditable linear model.

Section \ref{sec:Credit risk} describes the basics of credit risk modeling. In particular, we write the credit behavioral variables in terms of account type, balance, and outstanding credit, and show that the resulting daily paths carry a pronounced cyclical structure that monthly aggregation removes. This structure is what motivates the methodology of Section \ref{sec:FDA}, which introduces the preliminaries of functional data analysis and functional logistic regression, along with the relative time alignment that ensures the recurring events coincide across borrowers. Section \ref{sec:Experiments} reports our main findings. Section \ref{sec:conclusion} concludes and outlines directions for future work.

\section{Credit Risk Modeling}\label{sec:Credit risk}
In this section, we describe key elements of credit risk modeling and borrower credit risk behavior, along with a presentation of the data used in the analysis and subsequent experiments.

\subsection{Definition of Default}\label{sec:definition_of_default}

Our definition is based on an implementation of the European Banking Authority (EBA) guidelines of definition of default \citep{EBA_DoD_2016}.
Note that the following description is provided for illustrative purposes only; it should not be understood as a comprehensive or authoritative statement of the requirements, nor as guidance for achieving regulatory compliance.\\
\noindent A default is considered to have occurred when any of the following applies:
            \begin{itemize}
                \item The obligor is more than 90 days past due on any material credit obligation (90 DPD criterion).
                \item Borrower shows significant financial difficulty (general unlikeliness-to-pay case).
                \item Distressed restructuring leading to a Net Present Value (NPV) loss $\geq 1 \%$ triggers default.
                \item Bankruptcy, insolvency, or similar legal proceedings indicating the obligor cannot meet obligations.
                \item Credit write-off of the exposure by the institution.
                \item Sale of the credit obligation at a material loss due to credit deterioration.
                \item Other clear signs that the obligor will not repay in full without collateral realization.
            \end{itemize}

\subsection{Observational Times and Available Information}\label{sec:obstimes_and_variabletimes}
The term \textit{credit scoring} is used broadly in the literature, covering both the assessment made when a borrower applies for a loan \citep{XIA2017225} and the assessment of a borrower who already holds a credit obligation \citep{KWON2025125327}. The distinction is not always made explicit, even in the datasets commonly used for benchmarking: the German data describe applicants at the point of application, whereas the Taiwan and GMSC data summarize the behavior of borrowers who already hold an obligation \citep{KOZODOI20221083, OgundimuEmmanuelO.2019Podp}. The distinction is important, as different information is available for each task. Before turning to the variables themselves, we therefore explicitly state when the information used in this paper becomes available.

\begin{figure}[H]
  \centering
  \includegraphics[width=1\linewidth,height=6cm,keepaspectratio]{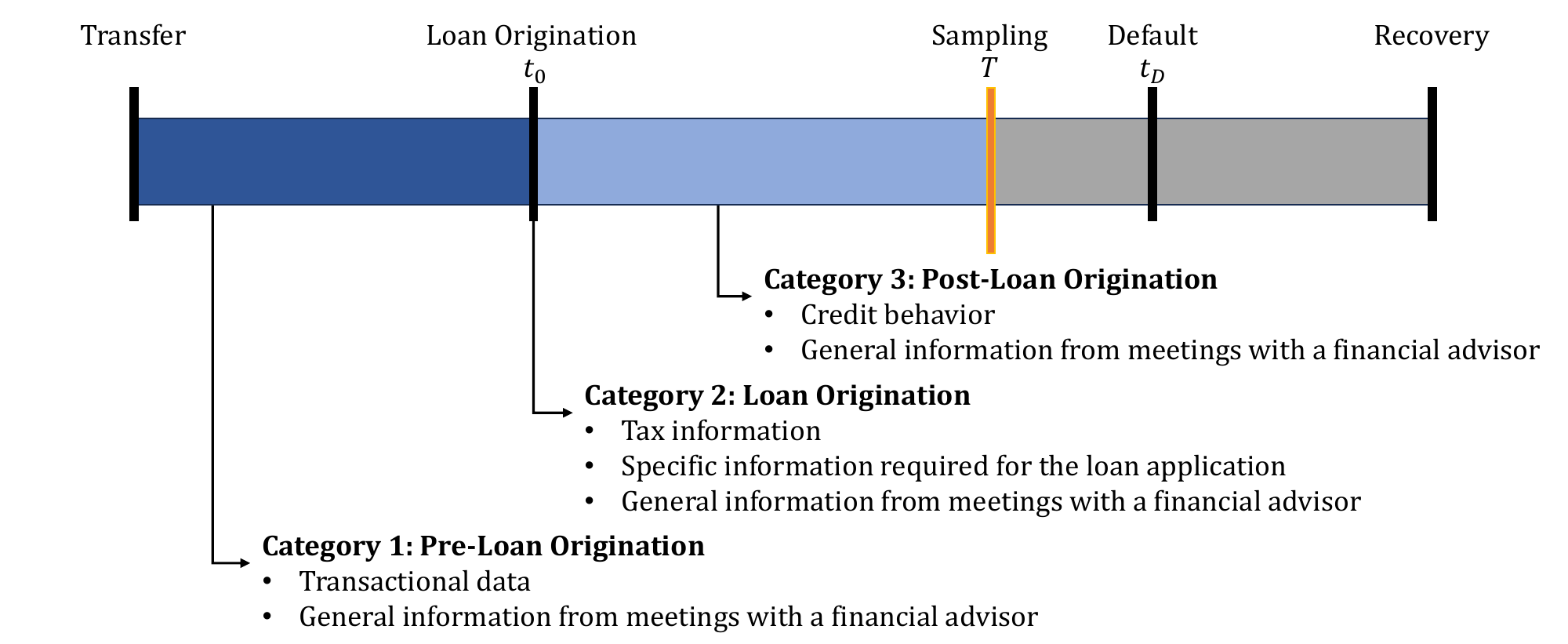}
  \caption{Visual illustration of when different information categories are available for retail borrowers.}
  \label{fig:variable_categories}
\end{figure}

Consider a retail borrower who transfers to a new financial institution and applies for a loan at time $t_0$. At loan application, the borrower submits the information required for a credit evaluation and, once the loan is granted, proceeds to manage the loan. While the loan is active, the institution must monitor the default risk, the need for impairments and if the borrower is in default. Each such monitoring point defines a sampling time $T$, at which the borrower is assessed. If the borrower subsequently meets one of the criteria of Section \ref{sec:definition_of_default}, default occurs at $t_D$, after which the exposure enters recovery. This timeline is illustrated in Figure\;\ref{fig:variable_categories}.
We divide the available information into three categories: \textit{pre-loan origination}, \textit{loan origination}, and \textit{post-loan origination}. Pre-loan origination is not described explicitly, as it is used to form the loan origination information.

\noindent\textbf{Category 2: At Loan Origination}\\
The credit evaluation conducted at loan origination relies on information that reflects the borrowers ability and willingness to repay, which the borrower is required to submit at $t_0$. This includes income statements, tax information, overviews of assets and liabilities, household budgets, and the purpose of the loan. The institution complements this self-reported data with external sources such as credit registries and credit bureaus. Together, these sources provide a comprehensive snapshot of the borrowers financial situation at $t_0$.

Loan origination information is the basis of most credit scoring models and is often regarded as the richest available data. However, it is typically only gathered at the time of loan application, since it is manually reported by the borrower. As a consequence, origination data becomes stale and may hold no information about the current financial status of the borrower, for instance in the case of a loss of income. Credit scoring models trained on origination data can therefore not detect post-origination deterioration unless the data is refreshed, which rarely occurs in practice, as it is manually intensive.\\

\noindent\textbf{Category 3: Post-Loan Origination}\\
After loan origination, the available information relies on what can be derived from transactional data, often referred to as credit behavior data. It is similar in nature to the pre-loan origination information, with the important addition that it also records how the borrower manages the granted obligation.

Where origination data is rich but static, post-loan origination data is updated continuously. It is therefore a natural basis for the ongoing monitoring of retail borrowers and is commonly the main input to retail IRB PD models. Although the underlying information is vast, it is in practice summarized in well known variables such as savings, delinquencies, available funds, utilization rate, and credit and debit turnover.\\

\noindent Due to the importance and availability of post-loan origination data, it forms the basis of our analysis. Below, we define the credit behavioral variables used in the analysis and examine the temporal structure of the underlying transactional data.

\subsection{Credit Behavioral Variables} \label{sec:ABCs}

There is, to the best of our knowledge, no established formal definition of credit behavioral variables. Hence, in this section we define credit behavioral variables in the following way.\vspace{0.4em}

\noindent For a given borrower $l$, observable at $t \in [t_{0,l},T_l]$, define
\begin{itemize}
    \item $A_{k,l}(t) \in \{1,2,\ldots,N\}$,  $N\in\mathbb{N}$ as the account type, for account $k$ of borrower $l$ at time $t$,\footnote{The number of categories of $A_{k,l}(t)$ can vary for each financial institution. We restrict the analysis to four account types, described in \ref{sec:account_types}.}
    \item $B_{k,l}(t) \in \mathbb{R}$ as the balance of account $k$ of borrower $l$ at time $t$,
    \item $C_{k,l}(t) \in \mathbb{R}^+$ as the size of the credit limit of account $k$ for borrower $l$ at time $t$.
\end{itemize}
Then a credit behavior variable is defined as any variable $X$ that can be written as a function of account type, balance, credit, or a combination of any of the variables.\vspace{0.4em}\\
\noindent Using the definition of credit behavior variables, we introduce three variables that are typically of interest in PD models. We will refer to them as available funds, utilization rate, and overdraft. They are given, respectively, as
\begin{align}
    \mathcal{L}_{k,l}(t) &:= B_{k,l}(t) + C_{k,l}(t), \label{eq:liquid}\\
    \mathcal{U}_{k,l}(t) &:= \boldsymbol{1}_{\left\{A_{k,l}(t)=2\right\}} \max\left( \frac{-B_{k,l}(t)}{C_{k,l}(t)},0\right),\label{eq:uddnyt}\\
    \mathcal{O}_{k,l}(t) &:=  \min\left(\mathcal{L}_{k,l}(t),0\right),\label{eq:overtrak}
\end{align}
where $A_{k,l}(t)=2$ refers to accounts with a line of credit attached. More information on account types can be found in \ref{sec:account_types}.\\

\noindent Each of the variables in equations \eqref{eq:liquid}-\eqref{eq:overtrak} has a clear interpretation, regardless of the type of account. This property allows us to aggregate information to the level of the borrower, by summing the accounts of each variable, e.g., for the available funds we write
\begin{align} \label{eq:aggr_custom}
    \mathcal{L}_{l}(t) &:= \sum_{k=1}^{n_l(t)}\mathcal{L}_{k,l}(t),
\end{align}
with $n_l(t)$ being the number of accounts that borrower $l$ owns at time $t$. The aggregated balance $B_l(t)$, credit $C_l(t)$, and overdraft $\mathcal{O}_l(t)$ are defined analogously as sums over the accounts of borrower $l$. At the household level, the sums are taken over all accounts held by the borrowers in the household. Aggregating variables to the borrower (household) level is important since borrowers (households) can freely move capital across accounts, creating the illusion of spending while capital never leaves the borrower (household).

The last two processes we will consider are only defined at aggregated levels, namely the gearing of the customer and the relative overdraft, defined respectively as
\begin{align}
    \mathcal{G}_l(t) &:= \frac{C_l(t)}{\mathcal{L}_{l}(t)},\label{eq:gear}\\
    \mathcal{R}_{l}(t) &:= \frac{\mathcal{O}_l(t)}{C_l(t)}.\label{eq:relovertrak}
\end{align}
Note that $\mathcal{O}_l(t)\leq 0$, and hence $\mathcal{R}_l(t)\leq 0$, so a larger overdraft corresponds to a more negative relative overdraft.

\subsection{Data}\label{sec:data}
The data used in this paper includes all transactions of retail borrowers of Nykredit, enriched with information regarding the borrowers accounts, the borrowers themselves, and a default flag. Using the descriptions in Section \ref{sec:obstimes_and_variabletimes} we have post-loan origination data for all borrowers.

When creating the default flag, say $Y$, we align with IRB regulations, restricting the  flag to a one year period. Thus, for some borrower $l$ at observation time $T_l$, if any of the criteria in Section \ref{sec:definition_of_default} are met within the next 12 months, the borrower is considered to be in default, i.e., $Y_l = 1$.

To ensure that sufficient information is available, the data is restricted to borrowers with a minimum of 12 months of loan history from a randomly assigned end-of-month sampling time $T$. Note that the restriction of 12 months of loan history does not violate any regulatory requirements, as long as the institution has separate models for the remaining borrowers with less than one year of post loan origination history.

We use a snapshot of the data from 2017-08-31 until 2019-12-31, which consists of 129,789 unique borrowers as the basis of the analysis and subsequent modeling. Furthermore, we enrich the model evaluation with out-of-time (OOT) dataset from 2021-01-31 until 2024-12-31, consisting of 177,338 borrowers.

In total, we have 307,127 borrowers. Each borrower has an average of 6 unique accounts and an average of 5255 transactions over the past year, including transactions between own accounts, translating to a daily average of approximately 14 transactions. At every transaction, the balance of the account is updated, meaning that the balance process jumps frequently.

\subsection{Properties of Account Type, Balance and Credit}\label{sec:properties_CBV}
A joint property of all three underlying processes is that they are observable at any time $t\in[t_0,T]$. Specifically, the following applies:

\begin{itemize}
    \item $A_{k,l}(t)$ can only change in the given data if a financial advisor manually changes it. This happens very rarely in practice, hence $A_{k,l}(t)$ will often be a constant category.
    \item $B_{k,l}(t)$ will change with every transaction on the account. Within transaction times, the value of the balance will remain constant. Thus, the balance process belongs to the C\`adl\`ag space with jumps at the times of transactions.
    \item $C_{k,l}(t)$ updates in accordance with the agreed upon credit obligation of the account. Usually, this entails that the size of the credit limit will decrease quarterly as installments are repaid, and increase when interest is drawn from the account. The size of the credit limit is thus a C\`adl\`ag process on $\mathbb{R}^+$.
\end{itemize}

\noindent While all three processes are continuous-time, we observed several practical issues with planned transactions during the initial data analysis. Planned transactions occur in batches between 2am and 4am. During this time span, we observe large variance in the balance process and potentially large technical overdrafts due to all outgoing transactions happening before all ingoing transactions. Furthermore, we have observed that most transactions can only occur during banking days, meaning that transactions during weekends or holidays are batched as planned transactions until the succeeding banking day.

To overcome the technical issues, and reduce dimensionality, we restrict all paths to being observed at the end of every banking day. The choice is in alignment with the definition of default in Section \ref{sec:definition_of_default}, where the 90 DPD criterion is evaluated at the end of the day. Furthermore, it is acceptable from a credit risk standpoint: a credit expert cares whether a borrower spends their monthly budget on the first banking day of the month but not when during the day.

\subsubsection{The Structure of Balance and Credit}
\noindent \textbf{Individual Paths}\\
To visualize individual balance and credit paths in the same figure, we plot the balance, $B_l(t)$, and the negative credit, $-C_l(t)$. In this setup, we expect the balance path (blue) to be above, but close to, the negative credit path (red), since the negative credit path can be viewed as the allowed negative balance for the borrower, i.e. if $B_l(t)< -C_l(t)$ the borrower is considered to have negative available funds.

\begin{figure}[H]
  \centering
  \includegraphics[width=1\linewidth,height=8cm,keepaspectratio]{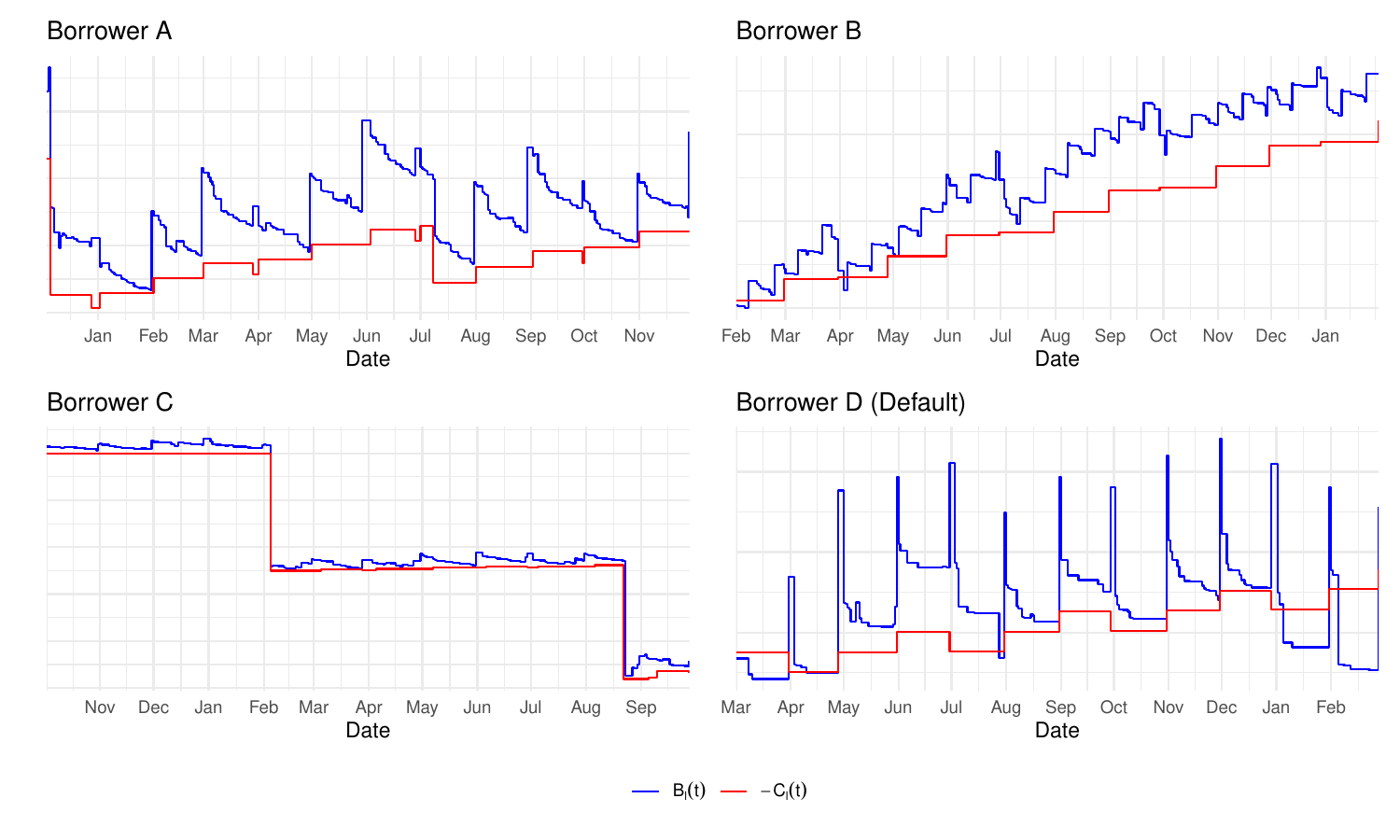}
  \caption{Illustrations of the balance and credit paths of individual borrowers. Borrowers A, B, and C are all considered normal borrowers, while borrower D defaults within the coming year. The Y-axis is deliberately hidden.}
  \label{fig:panel_of_customers}
\end{figure}

Four examples of the paths can be seen in Figure \ref{fig:panel_of_customers}. We find that the representation is a convenient way of viewing a borrower, providing a simple and easily interpretable overview of the borrowers finances.

\noindent From Figure \ref{fig:panel_of_customers} it is clear that the four borrowers behave very differently, highlighting the difficulty of modeling the credit risk of the borrowers. However, some underlying structure persists. We notice a positive jump in the balance every month due to salary payments, although borrower B appears to have multiple jumps, due to being paid bi-weekly. Moreover, the negative credit paths increase over time, as installments are being repaid. From borrower C, we note that the size of the credit obligation increases considerably two times during the observation period. The increase is due to being granted additional loans during the observation period, which is something to be aware of when modeling the credit risk of the borrowers. In particular, the balance of the customer would decrease greatly, possibly indicating risk, however, this is allowed as shown in the increase in the size of the credit obligations.

When considering the defaulting borrower D, we notice that it has large fixed expenses, since the balance decreases greatly on the first banking day, relative to the salary income. This creates a pattern that, when viewed at the end of the month seems healthy, as the borrower has available funds and has no pattern of decreasing savings. However, given the observations on banking days, it is clear that much of the income is absorbed by fixed expenses on the first banking day, making the customer vulnerable in case of unforeseen expenses. This more truthful representation of the borrower is one of the main motivations in this study, to consider not only end of month observations but the entire intra-monthly behavior. However, before presenting how to use the intra-month information, we will investigate the balance and credit paths aggregated across borrowers to understand the dynamics of the two processes.\\

\begin{figure}[H]
  \centering
  \includegraphics[width=1\linewidth,height=8cm,keepaspectratio]{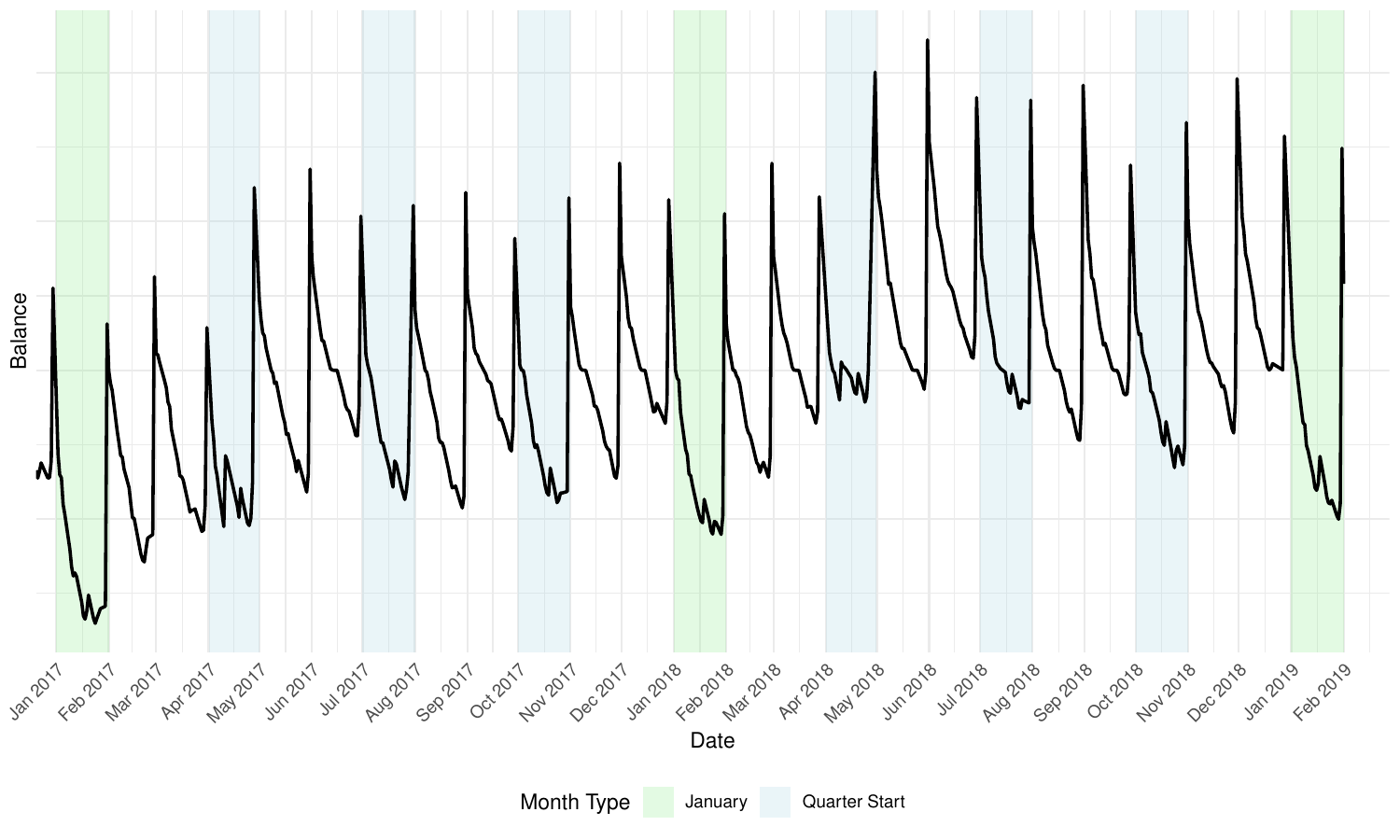}
  \caption{The median balance across all borrowers. The Y-axis is deliberately hidden.}
  \label{fig:medianbalance_dec2016_feb2019}
\end{figure}

\noindent\textbf{Balance}\\
To understand the dynamics of the balance paths, we will consider the median balance across all borrowers, as shown in Figure \ref{fig:medianbalance_dec2016_feb2019}. The median is chosen since the balance can vary greatly, with wealthy borrowers skewing averages.

From Figure \ref{fig:medianbalance_dec2016_feb2019} we notice several structures:
\begin{enumerate}
    \item \textbf{Spikes:} The median balance of the borrowers increases on the last banking day of the month. Likewise, the median balance drops sharply on the first day of the new month, before decreasing steadily throughout the month. This behavior results in a spike at the start of every month. We find that the spike arises from salary payments on the last banking day of the month and fixed payments on the first banking day of the next month.
    \item \textbf{Quarterly season:} The end-of-banking-month jump is smaller at the end of a quarter, resulting in a generally lower balance on the first month of each quarter. The first month of every quarter is highlighted on Figure \ref{fig:medianbalance_dec2016_feb2019} by either a blue or green color. This behavior is thought to arise from quarterly payments, such as installments on mortgages, insurance, and possibly external debt.
    \item  \textbf{Costly start to the new year:} Every January, the balance of the borrowers decreases more than in any other month, highlighted in green on Figure \ref{fig:medianbalance_dec2016_feb2019}. This behavior is partially expected from the quarterly seasonality, but the magnitude of the January drop exceeds what quarterly payments alone would explain. We attribute the excess to costs pooled during the Christmas holidays and advance payments for vacations later in the year.
    \item \textbf{Positive Trend:} Lastly, we observe a general increase in the balance of the borrowers. This trend is expected since the credit evaluation when granting a loan ensures that a borrower is financially stable and capable of increasing savings. Furthermore, the positive trend indicates a median wealth increase in the borrowers of the bank, possibly serving as a proxy for the economy. We will however not explore this further, in this paper.
\end{enumerate}

\noindent A further exploration of the daily change (first difference) of the balance for each borrower can be found in \ref{sec:first_diff_balance}.\\

\noindent\textbf{Credit}\\
When investigating the credit paths, we notice that the paths do not follow any clear cyclical pattern. However, we notice a slight positive trend in the credit paths, meaning that customers accumulate more debt as time increases. The property of increasing trend is surprising, as customers with loans should pay installments and thus decrease their credit. Upon further exploration, we find that being granted additional credit obligations offsets the payment of installments, as was the case for borrower C in Figure \ref{fig:panel_of_customers}. For a deeper analysis of the credit paths we refer to \ref{sec:appendix_credit_figures}.

\section{Functional Data Analysis}\label{sec:FDA}
As discussed in Section \ref{sec:properties_CBV} we restrict the data to end-of-banking days. This in turn leads to unequal and non-equidistant observations of the balance and credit paths, when compared across the random sampling time of borrowers. To avoid padding or truncation of the data, we suggest to employ the machinery of functional data analysis (FDA) and treat the non-equally spaced observations as realizations of a continuous-time process. This approach allows us to represent borrowers truthfully in the given setting.

\subsection{Functional Data Analysis and Logistic Regressions}\label{FDA_preliminaries}
In FDA, observations are viewed as realizations of random functions instead of finite-dimensional vectors \citep{FDA_Book_Ramsay_2005}. More precisely, given observations of the form $X(t_j)$,  $t_j\in\mathcal{T}$, $j=1,\dots,M$, FDA treats the collection $\{X(t_j)\}_{j=1}^M$ as evaluations of an underlying random trajectory $X(\cdot)$, rather than as $M$ separate measurements. Accordingly, $X(\cdot)$ is modeled as a random variable taking values in an infinite-dimensional function space.

The aim of this paper is to use FDA techniques for modeling a binary response $Y$ (default and not default) in terms of functional predictors constructed via \textit{credit behavioral variables}. We do this within the \textit{functional logistic regression} framework. Specifically, the components of the predictor $X=(X_1,\ldots,X_d):\mathcal{T}\to\mathbb{R}^d$ are assumed to take values in the space of square integrable functions $L^2(\mathcal T,\mathrm{d}s)$, in which $\mathrm{d}s$ represents the Lebesgue measure and $\mathcal{T}$ is a compact interval of $[0,\infty)$. We are interested in doing inference on the conditional probability
\[\pi(X):=\mathbb{P}(Y=1\rvert X),\]
which is further assumed to satisfy
\begin{equation}
\eta(X):=\mathrm{logit}(\pi(X))=\alpha+\sum_{j=1}^{d}\int_{\mathcal{T}}\beta_j(s) X_j(s)\,\mathrm{d}s,
    \label{eq:model}
\end{equation}
for some $\beta_1,\ldots,\beta_d\in L^2(\mathcal T,\mathrm{d}s)$ and where we used the notation $\mathrm{logit}(x)=\log(\frac{x}{1-x})$, $x\in(0,1)$. For ease of notation, the remainder of this section is presented for a single functional predictor ($d=1$), writing $X$ and $\beta$ for $X_1$ and $\beta_1$. The data extracted from the $i$th customer is represented as $(\{X_i(t_{i,j}):t_{i,j}\in\mathcal{T},j=1,\ldots,m_i\},Y_i),$ $i=1,\ldots,n$. As in standard logistic regression, we suppose that the data forms an i.i.d. sample across $i$.

Unlike the finite-dimensional setting, the parameter $\beta$ belongs to an infinite-dimensional space, making direct estimation infeasible without additional structure. A common strategy is to first approximate both the predictors $X$ and the coefficient $\beta$ through finite-dimensional basis expansions and then apply standard regression methods to estimate such approximations. More precisely, let $(\phi_k)_{k\geq 1}$ and $(\varphi_k)_{k\geq 1}$ be two collections of basis functions of $L^2(\mathcal T,\mathrm{d}s)$. Then each path of $X_i$ and $\beta$ may be represented as
\begin{equation}
X^{}_i(\cdot)=\sum_{k=1}^{\infty} c^{}_{ik}\phi_k(\cdot),
\qquad
\beta^{}(\cdot)=\sum_{k=1}^{\infty} b^{}_k\varphi_k(\cdot).
  \label{expansion}
\end{equation}
As a first step in the estimation procedure, one truncates the previous representations to a finite number of basis functions, i.e.,
\begin{equation}
X^{}_i(s)\approx\sum_{k=1}^{K} c^{}_{ik}\phi_k(s)=:X_{i,K}^{}(s);
\qquad
\beta^{}(s)\approx\sum_{k=1}^{L} b^{}_k\varphi_k(s)=:\beta^{}_L(s),
  \label{functional_approx}
\end{equation}
for some $K,L\in\mathbb{N}$. The estimation procedure is then divided into two steps: 1) estimate $c^{}_{ik}$, denoted $\hat{c}_{ik}$, and 2) given $\hat{c}^{}_{ik}$, estimate the coefficients $b^{}_k$.

\subsubsection{Estimation of $c_{ik}$}\label{sec:representation_functional}
The choice of basis $(\phi_k)_{k=1}^K$ in \eqref{functional_approx} determines which features of a borrowers trajectory the reconstruction $X_{i,K}$ can preserve. We use a spline basis, which we now define.

A spline of order $q$ is a piecewise polynomial of degree $q-1$ joined at knots $\xi_1\leq\cdots\leq\xi_{P+q}$ so that it has continuous derivatives up to order $q-2$. The B-spline functions form a convenient basis for this space, defined by the Cox--de Boor recursion \citep{deboor1972,cox1972}. Starting from
\begin{equation}\label{eq:bspline_base}
\phi_{k,1}(s) =
\begin{cases}
1, & \xi_k \leq s < \xi_{k+1},\\
0, & \text{otherwise,}
\end{cases}
\end{equation}
the order-$q$ functions ($q\geq 2$) are
\begin{equation}\label{eq:bspline_recursion}
\phi_{k,q}(s) =
\frac{s - \xi_k}{\xi_{k+q-1} - \xi_k}\, \phi_{k,q-1}(s)
+
\frac{\xi_{k+q} - s}{\xi_{k+q} - \xi_{k+1}}\, \phi_{k+1,q-1}(s),
\end{equation}
with terms having zero denominators set to zero. Writing $\phi_k\equiv\phi_{k,q}$, the resulting basis is nonnegative and forms a partition of unity, that is, $\sum_{k=1}^{K}\phi_k(s)=1$, and each $\phi_k$ is supported on only $q$ adjacent knot intervals.

This local support makes each coefficient $c_{ik}$ act on the curve only locally, so bursts of activity such as a large increase in the balance are captured without distorting the fit elsewhere. Splines also approximate any smooth function arbitrarily well as knots are added, without assuming a parametric form. For these reasons splines are standard for non-periodic functional data \citep{FDA_Book_Ramsay_2005}.

The coefficients $\mathbf{c}_i^{}=(c^{}_{i1},\ldots,c^{}_{iK})^T$ are estimated by fitting the observations $\{(t_{i,r}, X_i(t_{i,r}))\}_{r=1}^{m_i}$ to the spline basis via penalized least squares. Following \citet{FDA_Book_Ramsay_2005},
\begin{equation}\label{eq:penalized_ls}
\hat{\mathbf{c}}_i^{} = \argmin_{(c_{i1},\ldots,c_{iK})\in\mathbb{R}^K} \left\{ \sum_{r=1}^{m_i} \left( X_i^{}(t_{i,r}) - \sum_{k=1}^K c_{ik} \phi_k(t_{i,r}) \right)^2 + \lambda_X \int_{\mathcal{T}} \left( \sum_{k=1}^K c_{ik} \phi^{\prime\prime}_k(s)\right)^2 \mathrm{d}s \right\},
\end{equation}
where $\phi''(\cdot)$ denotes the second derivative and $\lambda_X \geq 0$ controls the trade-off between interpolation accuracy and smoothness of the reconstructed trajectory. Generally, larger values of $\lambda_X$ produce progressively smoother curves.

Once $\hat{\mathbf{c}}_i^{}$ has been obtained, the smooth functional representation is given by $\hat{X}_{i,K}^{}(t) = \sum_{k=1}^K \hat{c}^{}_{ik}\phi_k(t)$, which can be evaluated at any point $t\in\mathcal{T}$.

\subsubsection{Estimation of $b_k$}\label{sec:estimation_flm}
Given $\hat{\mathbf{c}}_i^{}$, the second stage estimates the slope function $\beta$. Substituting the truncated expansions in \eqref{functional_approx} into the linear predictor \eqref{eq:model} we get
\begin{equation}\label{eq:glm_approx}
\eta_i \;\approx\; \alpha +\sum_{l=1}^{L}\sum_{k=1}^{K} b_{l}\,\hat{c}_{ik}\int_\mathcal{T}\varphi_{l}(s)\phi_k(s)\, \mathrm{d}s.
\end{equation}
Defining the cross-product matrix $\mathbf{J}\in\mathbb{R}^{L\times K}$ with entries $J_{lk}=\int_{\mathcal{T}}\varphi_l(s)\phi_k(s)\,\mathrm{d}s$, the linear predictor becomes
\begin{equation}\label{eq:glm_matrix}
\eta_i \;\approx\; \alpha + \mathbf{b}^T \mathbf{J}\,\hat{\mathbf{c}}_i = \alpha + \tilde{\mathbf{x}}_i^T \mathbf{b},
\qquad \tilde{\mathbf{x}}_i := \mathbf{J}\,\hat{\mathbf{c}}_i \in \mathbb{R}^{L}.
\end{equation}
Thus the functional logistic regression reduces to an ordinary logistic regression of $Y_i$ on the $L$-dimensional transformed covariates $\tilde{\mathbf{x}}_i$, with parameters $\alpha$ and $\mathbf{b}=(b_1,\ldots,b_L)^T$. Writing $\pi_i=\pi(X_i)$, the log-likelihood is
\begin{equation}\label{eq:loglik}
\ell(\alpha,\mathbf{b})=\sum_{i=1}^{n}\Big[Y_i\log\pi_i+(1-Y_i)\log(1-\pi_i)\Big],\qquad
\pi_i=\frac{1}{1+\exp(-\alpha-\tilde{\mathbf{x}}_i^T\mathbf{b})}.
\end{equation}
Estimating $\mathbf{b}$ by plain maximum likelihood can lead to overfitting, since as the number of basis functions $L$ grows, the likelihood can be increased by fitting noise. When doing so, the unpenalized estimate of $\beta_L$ tends to fit idiosyncratic features of the training sample and produce a coefficient function that is neither stable nor meaningful in terms of inference. To control overfitting, we penalize the roughness of $\beta_L$. Using the integrated squared second derivative of $\beta_L$ as the roughness measure we get
\begin{equation}\label{eq:roughness_quadratic}
\int_{\mathcal{T}}\big(\beta_L''(s)\big)^2\,\mathrm{d}s
=\sum_{k=1}^{L}\sum_{l=1}^{L} b_k b_l \int_{\mathcal{T}}\varphi_k''(s)\varphi_l''(s)\,\mathrm{d}s
= \mathbf{b}^{T}\mathbf{S}\,\mathbf{b}, \quad S^{}_{kl} = \int_{\mathcal{T}} \varphi_k''(s)\,\varphi_l''(s)\,\mathrm{d}s.
\end{equation}
The estimator then maximizes the penalized log-likelihood
\begin{equation}\label{eq:penalized_loglik}
\ell_p(\alpha,\mathbf{b}) = \ell(\alpha,\mathbf{b}) - \frac{1}{2}\lambda_\beta\, \mathbf{b}^{T}\mathbf{S}^{}\mathbf{b}^{}.
\end{equation}
The maximizer $(\hat\alpha,\hat{\mathbf{b}})$ has no closed form and is obtained by penalized iteratively reweighted least squares, after which the estimated coefficient function is recovered as $\hat\beta_L(s)=\sum_{k=1}^{L}\hat b_k\varphi_k(s)$ and the fitted default probability for a new borrower follows from \eqref{eq:model}.

It remains to choose the smoothing parameter $\lambda_\beta$. Rather than fixing it by cross-validation, we estimate it by Restricted Maximum Likelihood (REML) \citep{wood2011fast}, which has been shown to provide stable estimates for generalized models with multiple smooth terms \citep{wood2017generalized}.

\subsubsection{Significance of the Coefficient Function}\label{sec:pvalues}
To assess if a functional predictor contributes to the model, we test whether its entire coefficient function is zero, $H_0:\beta\equiv 0$ against $H_1:\beta\not\equiv 0$. Since $\beta_L(s)=\sum_{k=1}^{L}b_k\varphi_k(s)$, this is a joint test of the basis coefficients $\mathbf{b}=\mathbf{0}$ and produces the single p-value reported for each functional variable in Table \ref{tab:func_summary}.

Because $\mathbf{b}$ is estimated under the roughness penalty in \eqref{eq:penalized_loglik}, an ordinary Wald test is anti-conservative, as the penalty shrinks $\hat{\mathbf{b}}$ and reduces its effective dimension below $L$. The reported p-value is therefore obtained from a test that accounts for this shrinkage, comparing a quadratic form in $\hat{\mathbf{b}}$ and its covariance against a reference distribution whose degrees of freedom are set by the effective degrees of freedom of the penalized term \citep{wood2013pvalues}. Such tests are provided by standard functional- and penalized-regression software \citep{goldsmith2011penalized, wood2017generalized}. The p-value is conditional on the smoothing parameter $\lambda_\beta$ and concerns the whole function; it does not localize where $\beta$ departs from zero, for which we rely on the bootstrap bands of Section \ref{sec:variable_select_functional}.

\subsection{Alignment via Relative Time}\label{sec:alignment}
In Section \ref{sec:properties_CBV} we have seen that a borrowers balance is mainly affected by two recurring events: salary is paid on the \emph{last} banking day of the month, and fixed obligations are debited on the \emph{first} banking day of the following month (Figure \ref{fig:medianbalance_dec2016_feb2019}). These events are tied to their ordinal position within the month rather than to a fixed calendar day.

Since we restrict every path to the end of each banking day, the number of monthly observations varies (between 19 and 23 in the data). This has two implications. First, months cannot be stacked into equal-length blocks, so a borrowers history has no natural fixed length. Second, and more importantly, the last banking day of the month does not sit at a fixed index: counting a fixed number of days back from a given sampling date lands on the end of the month for one borrower but mid-month for another whose relevant month had more banking days.

The first problem is solved using FDA and functional logistic regression, however, even when represented as continuous curves, the observations are misaligned, due to the second problem.
We propose using a common time axis on which the last banking day of the month is the same coordinate for every borrower. To construct it, consider a borrower sampled at date $T > 0$ with a lookback horizon of $M \in \mathbb{N}$ months, and label the months preceding $T$ as $m = 1, \ldots, M$, from the earliest to the most recent. For month $m$, let $d_1^{(m)}, \ldots, d_{k_m}^{(m)}$ denote its banking days in chronological order and $k_m$ their count.

It therefore only remains to reindex calendar time so that the last banking day of every month is mapped to a fixed value and only banking days receive a coordinate. Concretely, the relative time of the $j$-th banking day of month $m$ is defined as
\begin{equation}\label{eq:relative_time}
    \tau_j^{(m)} := \frac{j}{k_m} + (m - 1), \qquad j = 1, \ldots, k_m,
\end{equation}
the full construction is summarized in Algorithm~\ref{alg:relative_time}.

\begin{algorithm}[H]
\caption{Relative Time Construction}\label{alg:relative_time}
\begin{algorithmic}[1]
\Require Sampling date $T$, lookback horizon $M$ months
\Ensure Relative time $\tau_j^{(m)} \in (0, M]$ for each banking day in the observation window
\For{$m = 1, 2, \ldots, M$}
    \State Identify the $(M-m+1)$-th month preceding $T$
    \State Determine the banking days $d_1^{(m)}, \ldots, d_{k_m}^{(m)}$ and their count $k_m$
    \For{$j = 1, \ldots, k_m$}
        \State $\tau_j^{(m)} \leftarrow \dfrac{j}{k_m} + (m - 1)$
    \EndFor
\EndFor
\end{algorithmic}
\end{algorithm}

\noindent With the relative time construction in place, the credit behavioral variables produce trajectories that are directly comparable across borrowers and sampling dates, since the functional approach of Section \ref{FDA_preliminaries} already handles the varying number of banking days. By treating the observations $\{X(d_j^{(m)})\}$ as discrete evaluations of an underlying trajectory $X(\cdot)$, it is indifferent to whether a month contributes 20 or 23 points, as both reconstruct the same functional object.

\section{Experiments}\label{sec:Experiments}

To predict the probability of default, the 129,789 borrowers (described in Section \ref{sec:data}) are divided randomly into three datasets: training (73,152 borrowers), validation (31,352 borrowers) and test (25,285 borrowers).

We focus on the five variables defined in Section \ref{sec:ABCs}, namely available funds \eqref{eq:liquid}, utilization rate \eqref{eq:uddnyt}, overdraft \eqref{eq:overtrak}, gearing \eqref{eq:gear}, and relative overdraft \eqref{eq:relovertrak}.
When calculating the gearing of a customer, we floor the available funds, $\mathcal{L}$, to 1{,}000 DKK, mitigating extreme and missing values.
Furthermore, all variables are winsorized at their 2\% and 98\% percentiles to mitigate extreme values. The percentiles are not chosen point-wise, but rather at the sampling time of each individual borrower. They are computed on the training data only and applied unchanged to the validation, test, and OOT data. This is done before estimating the functional coefficients in \eqref{eq:penalized_ls}. \vspace{0.5em}

\noindent\textbf{Scalar Variables}\\
For the standard logistic regression and XGBoost models, each credit behavioral variable is reduced to a scalar by computing the arithmetic mean of its end-of-month values over a specified lookback horizon. Formally, let $X$ denote one of the variables $\mathcal{G}$, $\mathcal{R}$, $\mathcal{L}$, $\mathcal{O}$, or $\mathcal{U}$, defined in Section \ref{sec:ABCs}, then for a lookback of $M$ months, the scalar representation for borrower $l$ is
\begin{equation}\label{eq:scalar_variable}
    \bar{X}_l^{(M)} := \frac{1}{M} \sum_{m=1}^{M} X_l(\tau_{k_m}^{(m)}),
\end{equation}
where $\tau_{k_m}^{(m)} = m$ corresponds to the last banking day of month $m$, i.e.\ the end-of-month observation. For the special case $M = 0$, we define $\bar{X}_l^{(0)} := X_l(T)$, the value at the sampling date $T$.

We construct these averages for five lookback horizons: at observation ($M=0$), 3 months, 6 months, 9 months, and 12 months. Combined with five variables at two aggregation levels (borrower and household), this yields $50$ scalar features. Using horizons divisible by 3 is common practice, as it mitigates the quarterly seasonality documented in Section~\ref{sec:properties_CBV}.\vspace{0.5em}

\noindent\textbf{Functional Variables}\\
The functional variables are constructed from the end-of-bank-day observations of each of the chosen credit behavioral variables, mapped to the relative time domain $(0, M]$ as defined in Section~\ref{sec:alignment}. Each borrower $l$ thus provides a set of evaluations $\{(\tau_j^{(m)},\, X_l(\tau_j^{(m)}))\}$ for $j = 1,\ldots, k_m$ and $m = 1,\ldots, M$, where $X$ denotes one of the variables $\mathcal{G}$, $\mathcal{R}$, $\mathcal{L}$, $\mathcal{O}$, or $\mathcal{U}$.

From these discrete evaluations, we construct a smooth functional representation by expanding each trajectory in a spline basis, as described in Section \ref{sec:representation_functional}.

Several modeling choices arise in this construction:
\begin{enumerate}
    \item \textbf{Lookback horizon} ($M$). The number of months of history used to construct the functional variable. We consider $M \in \{3, 6, 9, 12\}$.
    \item \textbf{Basis type.} We only use cubic B-splines (placed at each observation $\tau_j^{(m)}$) and mean splines, which fully capture the variation in the data. Additional choices of basis representations have been examined, see \ref{sec:append_choice_of_basis}. We find that other representations do not change the model performance.
    \item \textbf{Smoothing parameter} ($\lambda$). For the cubic B-spline representation, we select $\lambda_X$ on the grid $\lambda \in \{0, 10^{-4}, 10^{-3}, 10^{-2}, 10^{-1},1\}$.
    \item \textbf{Aggregation level.} As for the scalar variables, the functional representation can be constructed at the borrower or household level.
\end{enumerate}

\subsection{Variable Selection}\label{sec:variabels}
Not all candidate variables carry predictive information, and many are highly correlated (e.g.\ gearing averaged over 6 and 9 months). Including all variables simultaneously would introduce multicollinearity in the logistic models and inflate variance without improving discrimination. We therefore apply variable selection separately for each model class, using the AUC on the validation set as the selection criterion throughout. For a further description of the AUC, and its relation to the Gini coefficient, we refer to \ref{sec:loss_funcions}.

For the standard and functional logistic regression, we will present the chosen variables and the sign of the beta coefficients, simply to ensure that the models are in alignment with basic credit intuition. A further exploration of the coefficients will not be made, as the goal is to predict and not to do inference.

\subsubsection{Standard Logistic Regression}\label{sec:variable_select_logistic}
In the standard logistic regression, we employ a greedy forward selection over the 50 scalar candidates defined in Section \ref{sec:Experiments}. At each step, the variable whose inclusion yields the largest AUC improvement on the validation dataset is added to the model. The procedure terminates when no remaining candidate improves the AUC by more than 0.05 percentage points.

The selected variables and their significance levels can be seen in Table \ref{tab:std_logistic_summary}.

\begin{table}[H]
\centering
\caption{Overview of the scalar variables used by the standard logistic regression}\label{tab:std_logistic_summary}
\begin{tabular}{lll}
\toprule
Variable      & Estimate  & p-value \\ \midrule
Intercept   & Negative      & $<0.001$              \\
Available funds (Household, $M=0$) & Negative & $<0.001$              \\
Gearing (Household, $M=6$)  & Positive      & $<0.001$              \\
Relative overdraft (Borrower, $M=12$)  & Negative     & $<0.001$             \\
\bottomrule
\end{tabular}
\end{table}

The interpretation of the three selected variables is in accordance with credit risk intuition.

\subsubsection{XGBoost}
The unrestricted XGBoost model receives all 50 scalar variables simultaneously and performs implicit variable selection through tree construction. Hyperparameters (maximum depth, learning rate, subsample ratio, and number of boosting rounds) are tuned via the validation set, optimizing for the AUC.

We acknowledge that grid search, in combination with cross-validation, is the standard procedure in the literature \citep{GUNNARSSON2021292,YU2009942,Alaraj2021,DUMITRESCU20221178}. However, we have tested our proposed method against variable and parameter selection on a 5-fold cross-validation and find no change in the resulting models, both for the standard logistic regression and the XGBoost. Thus, for ease of computation, and to align with the method proposed in Section \ref{sec:variable_select_functional} we use the training/validation dataset procedure.

\subsubsection{Functional Logistic Regression}\label{sec:variable_select_functional}
The procedure for the functional model mirrors the forward selection of the standard logistic regression: at each step, the functional variable that yields the largest AUC improvement is added to the model. The procedure terminates when no candidate improves the AUC by more than 0.05 percentage points. As described in Section \ref{sec:estimation_flm}, when training the functional logistic model, we need to specify the number of basis functions used for the beta function. Due to computational time, we always run the models using only $5$ basis functions, i.e., $L=5$.

The final selected variables can be seen in Table \ref{tab:func_summary}.

\begin{table}[H]
\centering
\caption{Overview of the functional variables used by the functional logistic regression}\label{tab:func_summary}
\begin{tabular}{ll}
\toprule
Variable                                                       & p-value \\ \midrule
Gearing (Household, $M=9$, cubic splines, $\lambda_X$=0.0001)       & $<0.001$ \\
Relative overdraft (Borrower, $M=12$, mean splines, $\lambda_X$=0)                  & $<0.001$ \\
Available funds (Household, $M=3$, cubic splines, $\lambda_X$=1)   & $<0.001$ \\
\bottomrule
\end{tabular}
\end{table}

\begin{figure}[H]
  \centering
  \includegraphics[width=.77\linewidth,height=13cm,keepaspectratio]{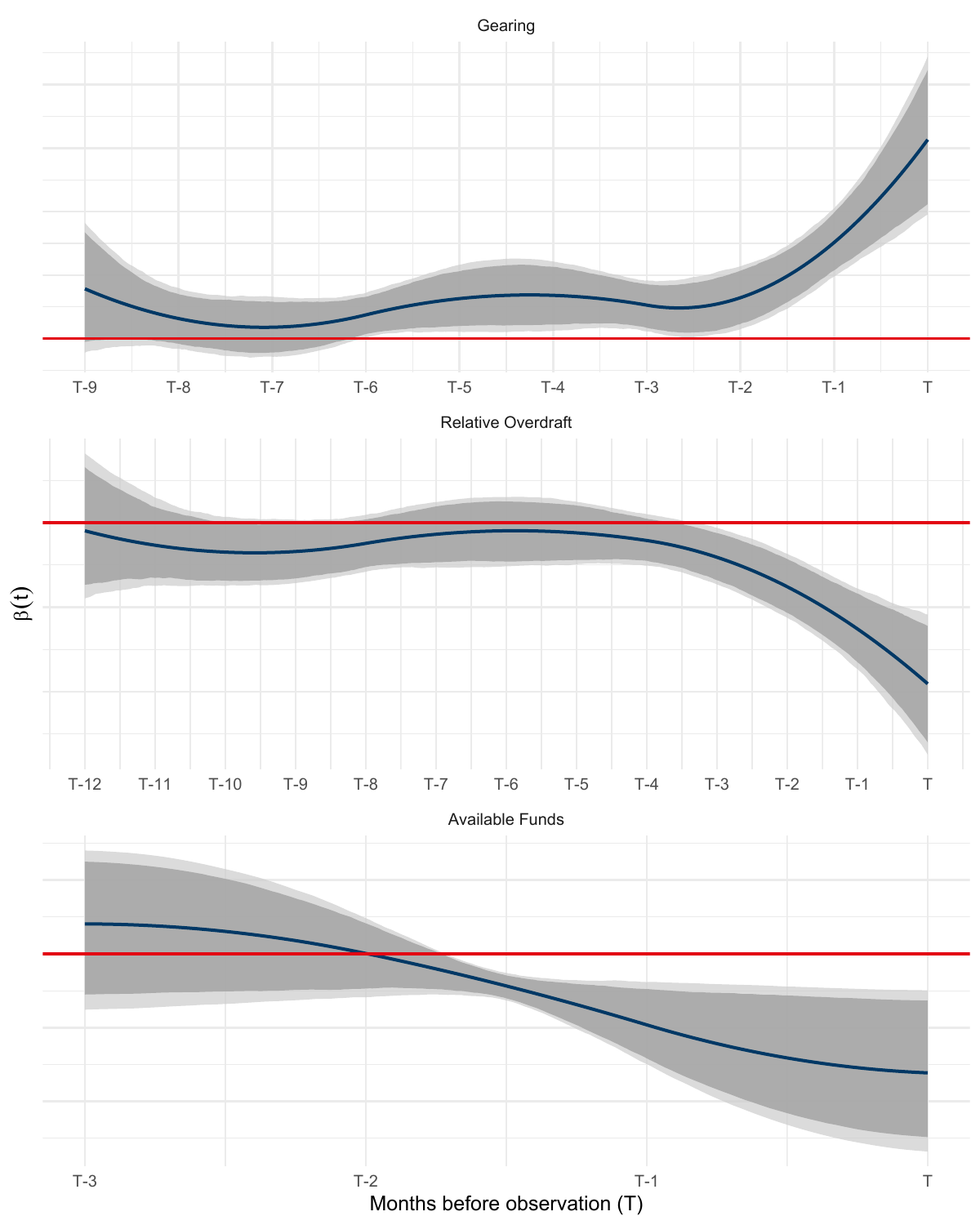}
  \caption{The $\hat{\beta}$ functions for each of the three functional variables}
  \label{fig:beta_functions}
\end{figure}

\noindent As noted in Section \ref{FDA_preliminaries}, the coefficients of the functional model are themselves functions and are therefore more exposed to overfitting than scalar coefficients. To confirm that the fitted model remains consistent with credit intuition, the estimated coefficient functions are shown in Figure \ref{fig:beta_functions}. In each panel the horizontal axis runs in months before the sampling date $T$. The shaded regions show pointwise $90\%$ and $95\%$ bootstrap confidence bands, constructed from 1{,}500 refits of the model on resamples of the training data drawn with replacement at the full training size, for more details see \citet{davison1997bootstrap} and \citet{goldsmith2011penalized}. In each refit the smoothing parameter $\lambda_\beta$ is held fixed at the value selected by REML on the full training data, rather than re-estimated on each resample. This isolates the variability of the estimated coefficient functions given the selected model from the variability of the smoothing selection itself.

From Figure \ref{fig:beta_functions} we see that the $\hat{\beta}$ functions for gearing and relative overdraft behave as expected, in terms of credit intuition, i.e., higher gearing and a larger overdraft is considered more risky.

The available funds coefficient is positive at the start of its three-month horizon, crosses zero around $T-2$, and becomes increasingly negative toward $T$. Read pointwise, the positive weight at the start is surprising, since credit intuition suggests that more available funds should lower default risk at every point in time, so one would expect a uniformly negative coefficient. However, in the functional model the variable enters only through the integral $\int \hat{\beta}(t)\,\mathcal{L}(t)\,\mathrm{d}t$, so what matters is not the sign of $\hat{\beta}$ at any single point but the weighted combination of the entire path of available funds. The shape of $\hat{\beta}$ encodes how the trajectory is aggregated, not an isolated snapshot.

Thus, with a large negative weight at $T$ and positive weight earlier in the window, the term contrasts the borrowers current available funds against their earlier level, and responds to the recent trend rather than to the level alone. Consider two borrowers who arrive at the same available funds at $T$: one whose funds have risen to that level and one whose funds have fallen to it. A snapshot model ($M=0$) assigns them the same risk, whereas this coefficient function assigns the borrower with the deteriorating trajectory the higher predicted default probability, since the drop from a higher earlier level enters through the positively weighted part of the window. This is the behavior that credit intuition would ask for, since a borrower whose liquidity is eroding is more concerning than one whose liquidity is building toward the same point. The large, significant negative weight at $T$ simultaneously preserves the expected level effect, so that, all else equal, higher current funds still lower the predicted risk.

\subsection{Model Comparison}
As is customary in credit risk, we use the AUC as the primary measure of discriminatory power \citep{LESSMANN2015124}. Furthermore, we evaluate the three models on the test dataset using four complementary metrics, namely the Brier score, log loss, weighted log loss, and focal loss. The reader may consult the appendix for a detailed explanation of these measures.

Table \ref{tab:model_performance} reports the out-of-sample performance of the three models.

\begin{table}[H]
\centering
\caption{Out-of-sample performance on the test set across five evaluation metrics. Bold indicates the best-performing model for each metric.}\label{tab:model_performance}
\begin{tabular}{llllll}
\toprule
Model              & AUC             & Brier             & LogLoss          & Weighted LogLoss  & Focal Loss \\ \midrule
Functional & \textbf{0.8996} & \textbf{0.007564} & \textbf{0.0363} & \textbf{1.7745}   & \textbf{0.00657} \\
Logistic   & 0.8902          & 0.007684          & 0.0374          & 1.8357            & 0.00687          \\
XGBoost    & 0.8940          & 0.007641          & \textbf{0.0363} & 1.7814            & 0.00664          \\
\bottomrule
\end{tabular}
\end{table}

The functional logistic regression achieves the highest AUC (0.8996), ahead of the standard logistic regression (0.8902) by 0.94 percentage points and the XGBoost model (0.8940) by 0.56 percentage points. The functional model thus attains a discriminatory power that is numerically higher than that of the XGBoost benchmark, although the difference is not statistically significant (Table \ref{tab:model_significance}), while remaining within the class of linear models, through a more faithful use of the data available for modeling.

Beyond ranking, we find the difference in performance to be comparable between the XGBoost model and the functional logistic regression, while both models outperform the standard logistic regression.

Given a moderate number of defaults in the test set, it is important to assess whether the observed AUC differences are statistically reliable. We apply paired DeLong tests \citep{delong1988} to compare the AUC of correlated ROC curves estimated on the same hold-out sample. We correct for multiple comparisons using the Holm-Bonferroni method across the two pre-specified pairwise comparisons.

\begin{table}[H]
\centering
\caption{Paired DeLong tests for AUC differences on the test set. $p$-values are two-sided and Holm--Bonferroni adjusted.}\label{tab:model_significance}
\begin{tabular}{lcccc}
\toprule
Comparison             & $\Delta$AUC       & Z-statistic       & Raw $p$-value        & Adjusted $p$-value\\ \midrule
Functional vs Logistic & 0.0094      & 2.569   & 0.0102   & 0.0204        \\
Functional vs XGBoost  & 0.0056      & 1.088   & 0.2765   & 0.2765                      \\
\bottomrule
\end{tabular}
\end{table}

The functional model significantly outperforms the standard logistic regression at the 5\% level, while the difference relative to XGBoost is not statistically significant. This pattern supports the conclusion that the functional logistic regression has an on-par performance with the XGBoost model. Section \ref{sec:oot} revisits this comparison on an out-of-time sample.

To isolate the contribution of the intra-monthly information, we refit the functional logistic regression using only the end-of-month observations of each functional variable. The model attains a test AUC of 0.8920, compared to 0.8996 for the daily functional model and 0.8902 for the standard logistic regression. Table \ref{tab:ablation_daily_monthend} shows that the month-end functional model is not significantly different from the standard logistic regression. Hence, the improvement over the standard logistic regression is driven mainly by the intra-monthly information rather than by the functional model itself.

\begin{table}[H]
\centering
\caption{Paired DeLong test for the AUC difference on the test set between the month-end functional model and the standard logistic regression. The $p$-value is two-sided.}\label{tab:ablation_daily_monthend}
\begin{tabular}{lccc}
\toprule
Comparison & $\Delta$AUC & Z-statistic & $p$-value \\ \midrule
Month-end functional vs Logistic & 0.0019 & 0.832 & 0.4052 \\
\bottomrule
\end{tabular}
\end{table}

\subsection{Performance on the Out-of-Time Sample}\label{sec:oot}
In practice, a PD model is developed on a fixed window and then used with little change for an extended period, so its performance on data from later periods is especially important. To assess the out-of-time performance, we use the OOT dataset, defined in Section \ref{sec:data}.

The OOT borrowers follow the same procedure as the development sample: the credit behavior variables are constructed from account type, balance, and outstanding credit (Section \ref{sec:ABCs}), mapped to the relative-time axis (Section \ref{sec:alignment}), and winsorized and aggregated, using the same fixed values. Hereafter, the three models of Section \ref{sec:variabels} are applied with their selected variables, basis choices, and smoothing parameters held fixed at the development-set values. This reflects how the models would be used in practice, where a model chosen on a development set is applied without change over a longer period.

Table \ref{tab:model_performance_oot} reports the out-of-time performance of the three models, and Table \ref{tab:model_significance_oot} the corresponding paired DeLong tests.

\begin{table}[H]
\centering
\caption{Out-of-time performance across five evaluation metrics. Bold indicates the best-performing model for each metric.}\label{tab:model_performance_oot}
\begin{tabular}{llllll}
\toprule
Model & AUC & Brier & LogLoss & Weighted LogLoss & Focal Loss \\ \midrule
Functional & \textbf{0.8901} & 0.006186 & 0.0309 & \textbf{1.6877} & \textbf{0.00573} \\
Logistic   & 0.8823 & 0.006278 & 0.0319 & 1.7264 & 0.00591 \\
XGBoost    & 0.8852 & \textbf{0.006146} & \textbf{0.0306} & 1.6905 & 0.00575 \\
\bottomrule
\end{tabular}
\end{table}

\begin{table}[H]
\centering
\caption{Paired DeLong tests for AUC differences on the out-of-time sample. $p$-values are two-sided and Holm--Bonferroni adjusted.}\label{tab:model_significance_oot}
\begin{tabular}{lcccc}
\toprule
Comparison & $\Delta$AUC & Z-statistic & Raw $p$-value & Adjusted $p$-value \\ \midrule
Functional vs Logistic & 0.0078 & 4.623 & $<0.0001$ & $<0.0001$ \\
Functional vs XGBoost  & 0.0049 & 1.806 & 0.0709 & 0.0709 \\
\bottomrule
\end{tabular}
\end{table}

The ranking observed in-sample is preserved out-of-time. All three AUC values are somewhat lower than on the in-time test set, as expected when models are applied to later periods, yet the functional model retains both the top rank and the largest margin over the logistic baseline.

The DeLong tests in Table \ref{tab:model_significance_oot} are consistent with the in-time results. The functional model exceeds the standard logistic regression by $0.0078$ AUC, which is significant at the 5\% level. Its advantage over XGBoost, $0.0049$ AUC, is not significant at the $5\%$ level. Overall, the out-of-time results indicate that the functional model improves on the standard logistic regression and remains at least on par with XGBoost when applied to later data.

\section{Conclusion}\label{sec:conclusion}
This paper shows that intra-monthly information improves the prediction of retail default. By using functional representations of three key credit behavior variables, the functional logistic regression significantly outperforms the standard logistic regression on both a test and OOT-dataset. Moreover, the functional logistic regression had on-par performance with an XGBoost model, indicating that the performance gap between standard logistic regression and machine learning models can be closed by using a richer representation of the available data. The similar performance is remarkable since the functional logistic regression is interpretable, allowing for the possibility of its use in a regulatory setup.

Functional data analysis provides a solid natural framework for representing credit behavioral data. The relative time construction of Section \ref{sec:alignment} places the recurring events at a common time, and the functional representation is indifferent to whether a month contributes 20 or 23 observations, so neither padding nor truncation is required.

This paper highlighted that credit behavioral variables used in practice are not separate pieces of information, but different functions of three underlying quantities: the type of each account, its balance, and the size of the credit limit. The representation opens several directions for future work. A natural extension of this paper would be to develop a functional model in which the quarterly cycle of credit behavior variables is captured in the model design. Furthermore, the representation naturally extends to other IRB models, in particular the Exposure at Default (EAD) model. Another large area of interest is behavioral monitoring, early warning systems, and potentially the generation of synthetic behavioral data. Lastly, we note that the representation can be used to learn what changes in borrower behavior over time, as a given borrower deteriorates towards default. It would be natural to expect broadly similar credit behavior at loan-origination, when correcting for loan-origination variables and the size of the credit obligation. After loan-origination, the borrowers defaulting will show signs of deterioration, creating potential flags for early warning and borrower monitoring systems.

\section*{Acknowledgments}
The authors would like to thank Jakob Lavröd for valuable discussions and feedback that helped improve the paper. Jonas Brunholm acknowledges support from Innovation Fund Denmark, grant number 4365-00039B.

\section*{Declaration of competing interest}
Jonas Brunholm is employed by Nykredit, which provided the data used in this study. The authors declare no other competing interests.

\section*{Data availability}
The data used in this study are proprietary to Nykredit and cannot be shared due to confidentiality and data protection restrictions. Code implementing the proposed methodology can be made available from the corresponding author upon reasonable request.

\printbibliography

\appendix

\section*{Appendix}
This appendix collects supporting material referenced in the main text. \ref{sec:account_types} describes the account types used in the analysis, \ref{sec:loss_funcions} defines the evaluation metrics, \ref{sec:first_diff_balance} and \ref{sec:appendix_credit_figures} present additional figures of the balance and credit paths, and \ref{sec:append_choice_of_basis} presents the functional model performance using different basis choices.

\section{Account types}\label{sec:account_types}

The analysis distinguishes four account types, defined below. Recall from Section \ref{sec:ABCs} that $A_{k,l}(t)$ denotes the type of account $k$ held by borrower $l$ at time $t$.

\begin{enumerate}[leftmargin=]
    \item[$A_{k,l}(t)=1$:] \textbf{Deposit accounts}. No line of credit or loan is attached. The borrower can freely withdraw from and deposit into the account.
    \item[$A_{k,l}(t)=2$:] \textbf{Credit accounts}. A line of credit is attached. The balance of the account can vary freely, as for the deposit accounts, however maximal allowed negative balance is $-C_{k,l}(t)$ and not 0.
    \item[$A_{k,l}(t)=3$:] \textbf{Loan accounts}. A loan is attached. The balance of the account cannot vary freely, as the only change in balance is due to paying installments. The due amount of the loan is $C_{k,l}(t)$, which is decreased when installments are paid.
    \item[$A_{k,l}(t)=4$:] \textbf{Investment accounts}. No line of credit or loan is attached. The balance of the account, $B_{k,l}(t)$, is the value of the financial assets. The borrower can freely buy or sell financial assets, changing the balance of the account.
\end{enumerate}

All other types of accounts, such as pension accounts or temporary accounts, are usually not of interest when modeling borrowers and are thus not defined.

\section{Loss Functions}\label{sec:loss_funcions}
In credit risk modeling, both institutions and regulators prioritize a model's ability to rank customers by risk. The relative ranking determines rating class assignments, which in turn drive pricing of interest rates and fees. This focus on ordering, rather than binary classification, requires loss functions that evaluate predicted probabilities without imposing a decision threshold.

We evaluate model performance using five metrics. The AUC captures discriminatory power, while the remaining four assess calibration and sensitivity to class imbalance. Let $y_i \in \{0,1\}$ denote the observed default indicator, $\hat{p}_i \in (0,1)$ the predicted probability of default for observation $i$, and $N$ the total number of observations.

\subsection{AUC (Area Under the ROC Curve)}
To measure the ranking power of a model we use the AUC, which can be expressed through the concordance of prediction pairs as
\begin{align*}
    \text{AUC}(\hat{p},y) &= \frac{C+\tfrac{1}{2}T}{C+D+T},\\
    C = \sum_{i<j}\mathbf{1}_{\{(y_i-y_j)(\hat{p}_i-\hat{p}_j)>0\}}&,\quad D = \sum_{i<j}\mathbf{1}_{\{(y_i-y_j)(\hat{p}_i-\hat{p}_j)<0\}}, \quad T = \sum_{i<j}\mathbf{1}_{\{y_i\neq y_j,\ \hat{p}_i=\hat{p}_j\}}
\end{align*}
where $C$ is the number of concordant pairs, $D$ is the number of discordant pairs, and $T$ is the number of pairs with tied predictions, all among pairs with different outcomes. Hence $C+D+T$ equals the number of defaults times the number of non-defaults.

The AUC is bounded between $0$ and $1$, with $0$ being a perfect negative ranking and $1$ being a perfect ranking. Furthermore, an AUC of $0.5$ means no ranking power. Thus, the AUC measures how well the model separates the predictions, but not by how much.

Note that the Gini coefficient (Somers' D), which is also commonly reported in credit risk, is a linear rescaling of the AUC, $\text{Gini} = 2\cdot \text{AUC}-1$, so the two induce the same ranking of models.

\subsection{Brier Score}
The Brier score measures the mean squared deviation between predicted probabilities and observed outcomes:
\begin{equation}
    \text{BS} = \frac{1}{N} \sum_{i=1}^{N} (\hat{p}_i - y_i)^2
\end{equation}
It is a strictly proper scoring rule that simultaneously penalizes poor calibration and poor discrimination. Unlike the AUC, the Brier score is sensitive to the level of the predicted probabilities, not only their ordering.

\subsection{Log Loss}
The log loss measures the negative average log-likelihood of the predictions:
\begin{equation}
    \text{LL} = -\frac{1}{N} \sum_{i=1}^{N} \left[ y_i \log(\hat{p}_i) + (1 - y_i) \log(1 - \hat{p}_i) \right]
\end{equation}
It penalizes confident but incorrect predictions heavily due to the logarithmic transformation. A model predicting $\hat{p}_i$ close to $0$ for an observation where $y_i = 1$ incurs a very large loss.

\subsection{Weighted Log Loss}
Given the class imbalance in our sample, we employ a weighted log loss that assigns weight $w_1$ to defaulting observations:
\begin{equation}
    \text{WLL} = -\frac{1}{\sum_{i=1}^{N} w_i} \sum_{i=1}^{N} w_i \left[ y_i \log(\hat{p}_i) + (1 - y_i) \log(1 - \hat{p}_i) \right]
\end{equation}
where $w_i = w_1$ if $y_i = 1$ and $w_i = 1$ otherwise. We set $w_1$ to be approximately equal to the inverse of the observed default rate. This ensures that the loss contribution from defaults is comparable in magnitude to the contribution from non-defaults, preventing the metric from being dominated by the majority class.

\subsection{Focal Loss}
The focal loss, introduced by \citet{lin2017focal}, down-weights well-classified observations and focuses on hard-to-classify cases:
\begin{equation}
    \text{FL} = -\frac{1}{N} \sum_{i=1}^{N} \alpha_i (1 - p_{t,i})^\gamma \log(p_{t,i})
\end{equation}
where $p_{t,i} = \hat{p}_i$ if $y_i = 1$ and $p_{t,i} = 1 - \hat{p}_i$ otherwise, and $\alpha_i = \alpha$ if $y_i = 1$ and $\alpha_i = 1 - \alpha$ otherwise. We use $\gamma = 2$ and $\alpha = 0.25$. The modulating factor $(1-p_{t,i})^\gamma$ reduces the loss contribution from observations that are already well-classified, effectively concentrating the evaluation on borderline cases where models differ most.

\section{First Difference of Balance Paths}\label{sec:first_diff_balance}

\begin{figure}[H]
  \centering
  \includegraphics[width=1\linewidth,height=10cm,keepaspectratio]{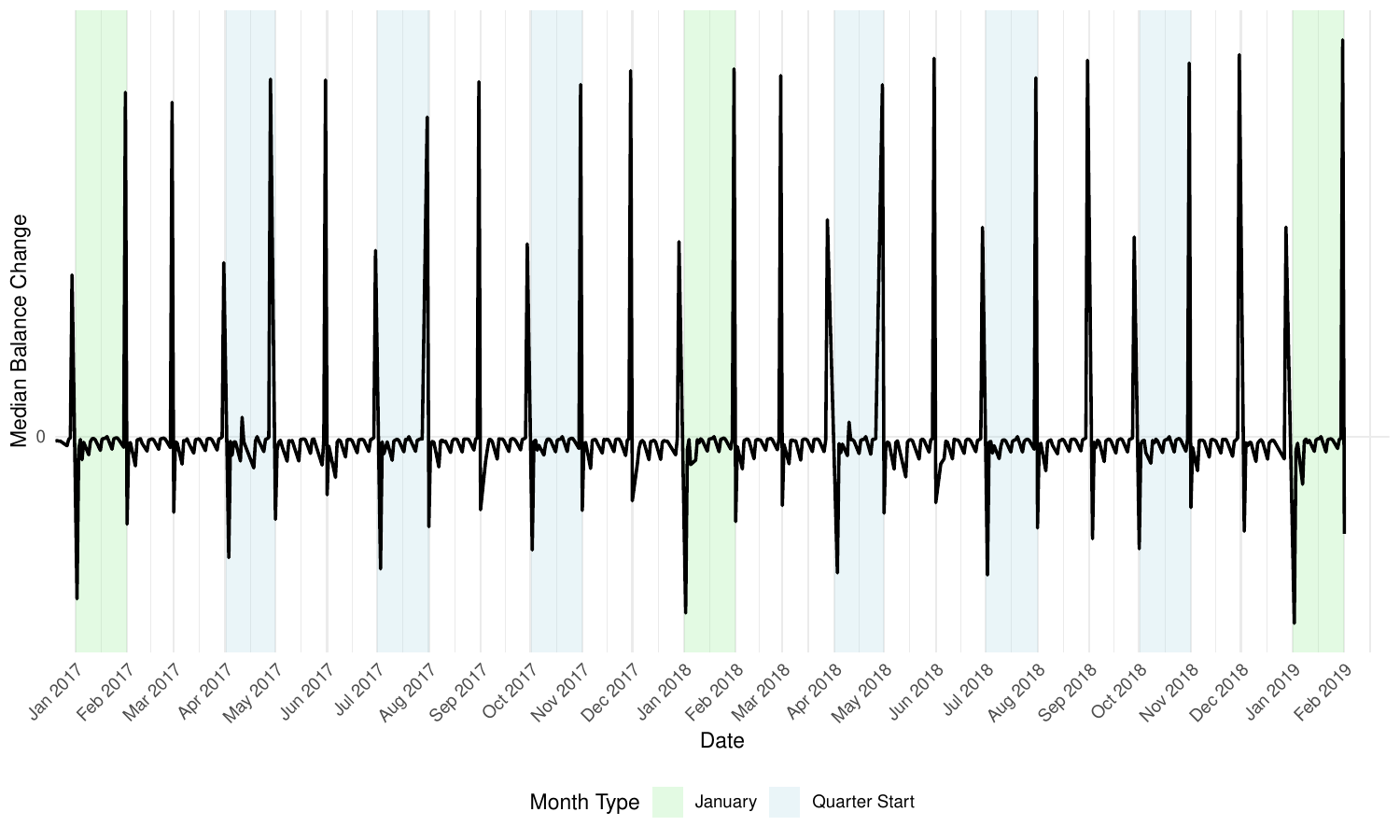}
  \caption{The median customer-specific daily lags of the balance}
  \label{fig:median_balancechange_dec2016_feb2019}
\end{figure}

\noindent An issue with Figure \ref{fig:medianbalance_dec2016_feb2019} is that the median across borrowers does not have to match any single borrower, since the median at each date is a cross-sectional summary, and it can be a different borrower from one date to the next. The level differences and the general upward trend can take over the plot and hide the within-borrower changes, which are what we are interested in. To separate the two, we take the daily lag (first difference) of the balance for each borrower, and then take the median of these lags across all borrowers, as shown in Figure \ref{fig:median_balancechange_dec2016_feb2019}. Taking the difference removes the level of each borrower, so the median lag tells us the typical daily change in balance instead of the typical balance. We note that this median of lags is still a cross-sectional summary and does not trace the path of one borrower either, but it does let us see the day-to-day movements more clearly.

When investigating Figure \ref{fig:median_balancechange_dec2016_feb2019} we see the same pattern as in Figure \ref{fig:medianbalance_dec2016_feb2019}, only expressed in lags: the same monthly spikes, quarterly pattern, and yearly effects reappear as jumps in the daily changes. However, the positive trend is naturally gone.

\section{Figure of The Credit Paths}\label{sec:appendix_credit_figures}

\begin{figure}[H]
  \centering
  \includegraphics[width=1\linewidth,height=10cm,keepaspectratio]{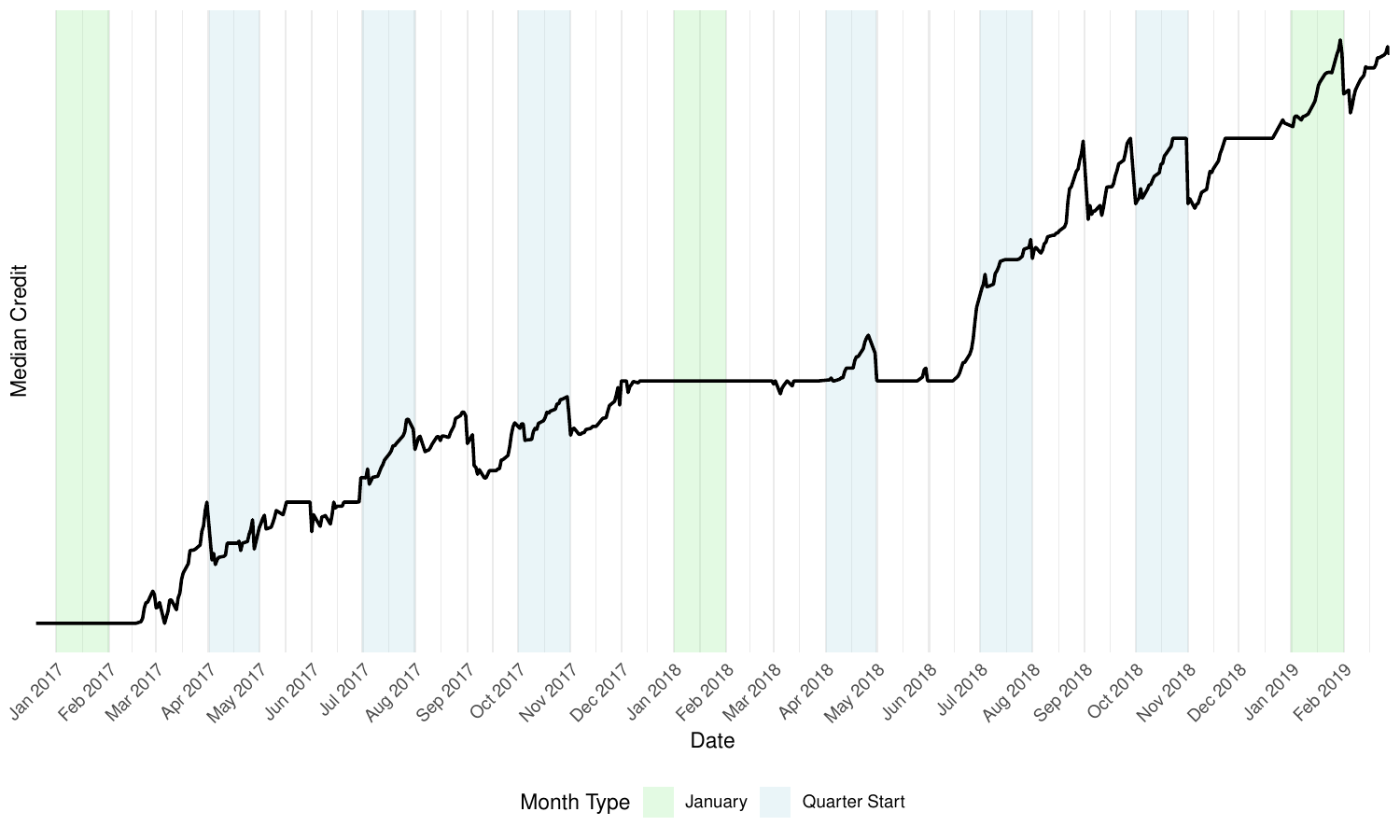}
  \caption{The median credit across all borrowers}
  \label{fig:figure_mediancredit_dec2016_feb2019}
\end{figure}

\begin{figure}[H]
    \centering

    \begin{subfigure}{\linewidth}
        \centering
        \includegraphics[width=\linewidth,height=10cm,keepaspectratio]{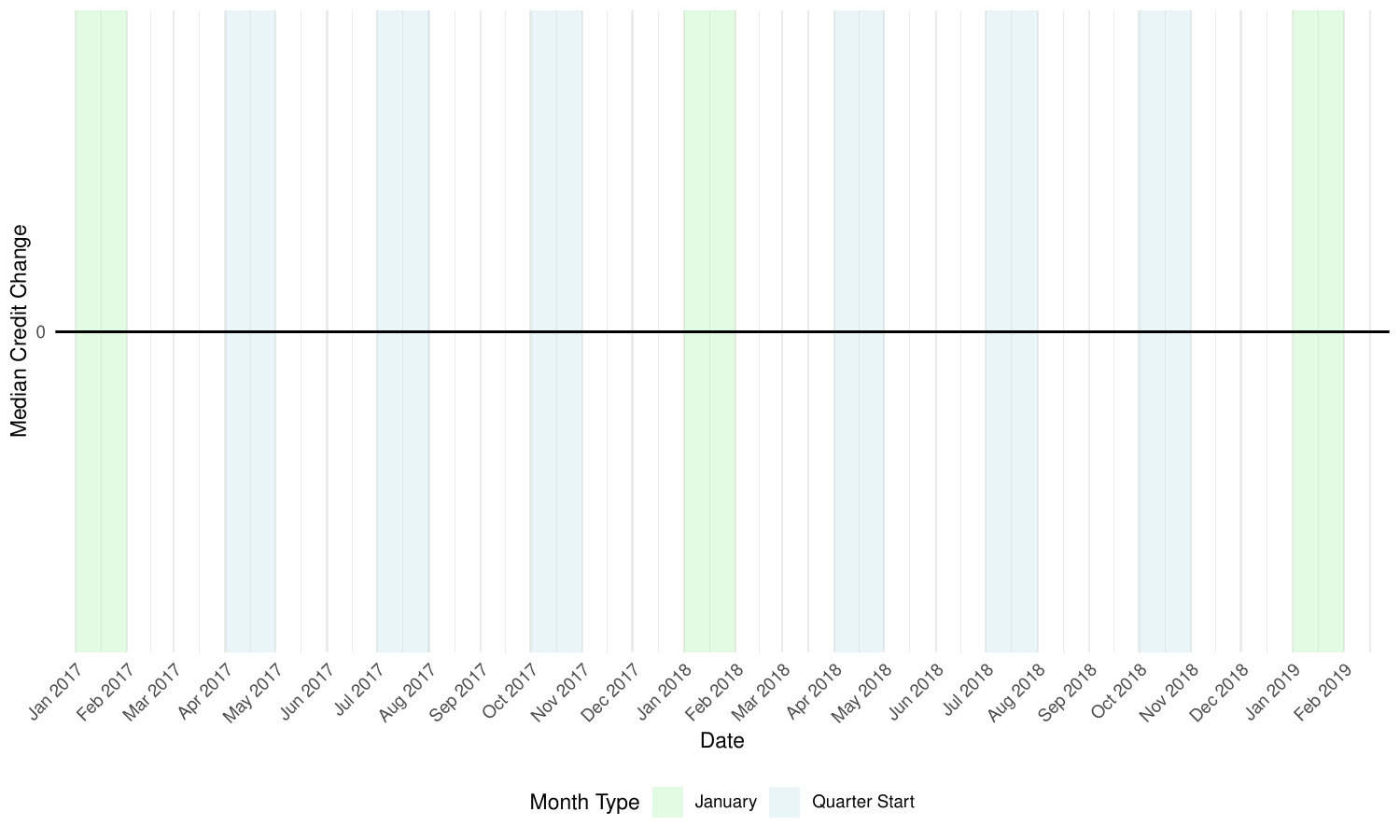}
        \caption{Median credit change across all borrowers}
        \label{fig:median_all}
    \end{subfigure}

    \vspace{0.5cm}

    \begin{subfigure}{\linewidth}
        \centering
        \includegraphics[width=\linewidth,height=10cm,keepaspectratio]{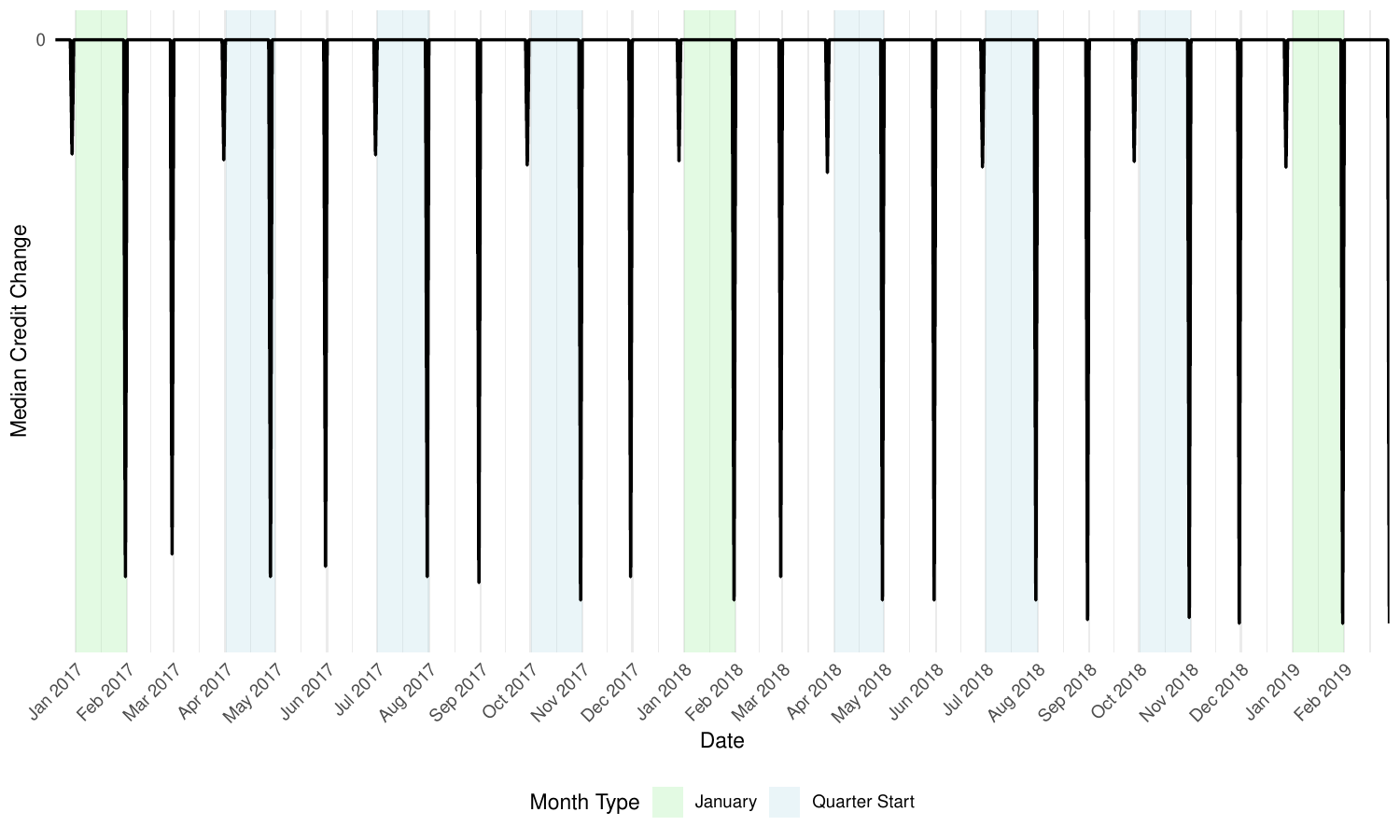}
        \caption{Median credit change across loan accounts}
        \label{fig:median_loans}
    \end{subfigure}

    \caption{Median credit change}
    \label{fig:combined_median_credit}
\end{figure}

This section collects the credit-path figures referenced in Section \ref{sec:properties_CBV}. Figure \ref{fig:figure_mediancredit_dec2016_feb2019} shows the median credit across all borrowers, while Figure \ref{fig:combined_median_credit} shows the median credit change, both overall and restricted to loan accounts.

From Figure \ref{fig:figure_mediancredit_dec2016_feb2019} we see that the median credit increases over the observation window. The rise is irregular rather than smooth, with stretches of little change, such as late 2017 to early 2018, interrupted by clearer upward steps, most visibly around mid-2018. In contrast to the balance in Figure \ref{fig:medianbalance_dec2016_feb2019}, the credit shows no pronounced monthly or quarterly cycle, and the January and quarter-start shading do not line up with any repeating pattern. The upward drift is in line with borrowers taking on additional credit obligations over time, which offsets the repayment of installments.

Figure \ref{fig:combined_median_credit} shows the daily change (first difference) of the credit, which removes the level and makes the day-to-day movements easier to see. Panel (a) aggregates across all accounts, where the median change is zero, since the credit limit is constant between scheduled events and only moves when credit is drawn or repaid. Panel (b) restricts attention to loan accounts, where the quarterly installment structure is more visible as recurring downward movements, reflecting the scheduled reduction of the outstanding credit as installments are paid.

\section{Other Choices of Basis}\label{sec:append_choice_of_basis}

Table \ref{tab:model_performance_basis} shows the resulting models, after forward selection, when using a different basis representation than the B-spline basis. Fourier denotes a Fourier basis, Local polynomial denotes local polynomial smoothing, and Mixed denotes forward selection where each candidate variable may be represented by either the Fourier basis or local polynomial smoothing. The mixed selection chose only Fourier-represented variables and therefore coincides with the Fourier model. We generally see no remarkable change in the performance of the models.

\begin{table}[H]
\centering
\caption{Performance across five evaluation metrics for alternative basis representations. Bold indicates the best-performing model for each metric.}\label{tab:model_performance_basis}
\begin{tabular}{llllll}
\toprule
Model & AUC & Brier & LogLoss & Weighted LogLoss & Focal Loss \\ \midrule
Benchmark (splines) & \textbf{0.8996} & \textbf{0.007564} & \textbf{0.0363} & \textbf{1.7745} & \textbf{0.00657} \\
Fourier             & 0.8989 & 0.007570 & 0.0364 & 1.7805 & 0.00660 \\
Local polynomial    & 0.8984 & 0.007565 & 0.0364 & 1.7803 & 0.00660 \\
Mixed               & 0.8989 & 0.007570 & 0.0364 & 1.7805 & 0.00660 \\
\bottomrule
\end{tabular}
\end{table}

\end{document}